\documentclass[aps,pra,a4paper,floatfix,showpacs,notitlepage,twocolumn]{revtex4-1} 
\usepackage{amsmath}
\usepackage{amsfonts}
\usepackage{amssymb}
\usepackage{graphicx}
\graphicspath{{./Figures/}}
\usepackage{epsfig}
\usepackage{epstopdf}
\usepackage{xcolor}
\usepackage{empheq}
\usepackage{cancel}
\usepackage{braket}
\begin{document}

\title{Nonlinear quantum Langevin equations for bosonic modes in solid-state systems}
\author{Juuso Manninen} 
\affiliation{Department of Applied Physics, Low Temperature Laboratory, 
               Aalto University, PO Box 15100, FI-00076 AALTO, Finland}
\author{Souvik Agasti} 
\affiliation{Department of Physics and
  Nanoscience Center, University of Jyvaskyla, P.O. Box 35 (YFL), FI-40014
  University of Jyvaskyla, Finland}
\author{Francesco Massel}
\email[]{francesco.p.massel@jyu.fi} \affiliation{Department of Physics and
  Nanoscience Center, University of Jyvaskyla, P.O. Box 35 (YFL), FI-40014
  University of Jyvaskyla, Finland}

\begin{abstract}
  Based on the experimental evidence that impurities contribute to the
  dissipation properties of solid-state open quantum systems, we provide here a
  description in terms of nonlinear quantum Langevin equations of the role
  played by two-level systems in the dynamics of a bosonic degree of freedom.
  Our starting point is represented by the description of the system/environment
  coupling in terms of coupling to two separate reservoirs, modelling the
  interaction with external bosonic modes and two level systems,
  respectively. Furthermore, we show how this model represents a specific
  example of a class of open quantum systems that can be described by nonlinear
  quantum Langevin equations. Our analysis offers a potential explanation of the
  parametric effects recently observed in circuit-QED cavity optomechanics
  experiments.
\end{abstract} 
\maketitle

The dynamics of open quantum systems --i.e. quantum systems that can be
described as separate entities from their surrounding environment while being
somehow coupled to it-- is arguably one of the most fundamental problems in
quantum mechanics, encompassing concepts such as the \textit{measurement
  paradox} \cite{Leggett:2005bf}, and the boundary between quantum and classical
physics \cite{ZUREK:1991td}. On general grounds, the interaction between a
quantum system and its environment represents an important aspect of the physics
of condensed matter and complex systems, which has been the focus of extensive
analysis \cite{Leggett:1987wk,Breuer:2007wp,Breuer:2015wd}, with repercussions
in contexts ranging from the energy transport in photosynthetic complexes
\cite{Ishizaki:2010cd} to the physics of ultracold gases
\cite{Leskinen:2010cy,Massel:2013ge,Visuri:2013vs}.

In the description of these systems the inclusion of the role played by coupling
to an external environment is necessary, if only because the system has to be
coupled to an external measurement apparatus which, from the quantum-dynamical
perspective of the system, represents a source of noise and dissipation.  At the
same time the manipulation of open quantum systems has recently led to the
possibility of preparing and detecting quantum states of matter and radiation
\cite{Anonymous:dwBv17Gu,Wiseman:2010vw}, paving the way for the definition of a
new paradigm of \textit{quantum technology} which represents an important field
for applications ranging from secure (quantum) communication
\cite{Nielsen:2000kz} to sensing of electromagnetic fields \cite{Clerk:2010dh}
and to the detection of gravitational waves \cite{Abbott:2016ki}. This prospect
of technological application of quantum mechanics is rooted in the relatively
recent development of fabrication techniques at the nanoscale, in particular
nanomechanical resonators, superconducting qubits and, more in general, circuit
quantum electrodynamics (QED) setups
\cite{Wallraff:2004dy,Sillanpaa:2007ig,Majer:2007em,Schoelkopf:2008cs} where the
characteristic scales involved in the dynamics of these systems naturally lead
to the study of the quantum properties in the presence of coupling to an
environment.

Within this framework, it has recently been observed that this coupling can
represent an important resource leading to the notion of \textit{reservoir
  engineering} \cite{Poyatos:1996ej}. This concept corresponds to the idea
that, by manipulating the properties of the environment coupled to a given
quantum system or even the nature of the system environment coupling itself, it
is possible to generate specific (quantum) states for the system. Prominent
examples are represented by the recent achievements in the field of cavity
optomechanics, where ground state cooling \cite{Teufel:2011jga} and squeezing
below the standard quantum limit (SQL) \cite{Wollman:2015gx,Pirkkalainen:2015ki, Lecocq:2015dk}, along
with nearly quantum limited amplification \cite{Massel:2011ca,
  OckeloenKorppi:2016ke} and nonreciprocal photon transmission
\cite{Metelmann:2015gb} have been achieved by introducing a specific (Gaussian)
state for the reservoir. While these examples correspond to inducing a specific
state for the system by manipulating the state of the reservoir, in
Refs. \cite{Mirrahimi:2014js,Leghtas:2015kd} it is shown that, by designing a
specific nonlinear coupling between system and environment: it is possible to
protect certain quantum states (cat states) against decoherence.

If the coupling between the system and the environment is described by a linear
Hamiltonian, the effects of noise and dissipation on the dynamics of the system
can be described in terms of linear quantum Langevin equations (QLEs)
\cite{Wiseman:2010vw}.  These equations represent an extension to the quantum
regime of the classical Langevin equations and, in analogy to their classical
counterpart, include in the description of the dynamics of the system the role
played by the environment, including dissipative and noise effects. However the
case of a linear system/environment coupling is not the most general situation
that can arise: for instance, for nanomechanical resonators
\cite{Zolfagharkhani:2005ds,Arcizet:2009ix,Eichler:2011fib,Suh:2012ic,Singh:2016jj}
and for circuit QED setups
\cite{Simmonds:2004ga,Martinis:2005kwa,Ashhab:2006ed,OConnell:2008jt,Gao:2008ks,Neill:2013ir},
the experimental evidence of nonlinear phenomena related to the coupling between
system and environment has emerged and, more importantly for our analysis, the
relevance of impurities in this phenomenon has been discussed. In both
setups, it has been shown that the impurities, naturally arising in the material
composing the devices, its supports and/or substrate, represent a source of
dissipation. These defects can be modelled in terms of TLSs. The reason behind
the possibility of modelling impurities in these terms is represented by the
fact that each impurity can be construed as quantum systems which exhibit two
local energy minima. For instance as a charged impurity that can hop between two
defects in the crystal structure, or a dangling bond with two possible configurations.

More specifically, these TLSs exist primarily due to the disordered potential
landscape of amorphous materials -- e.g. in surface oxides of thin-film circuit
electrodes \cite{Gao:2008ks}, in the tunnel barrier of Josephson junctions
\cite{Simmonds:2004ga}, and at disordered interfaces
\cite{Phillips:1987ge,Quintana:2014jp} -- coupling with the bosonic degrees of
freedom of the system, either through a purely electromagnetic interaction
(optical and circuit QED setups) or a phononic one in the context of
nanomechanical systems \cite{Kleiman:1987hm}.

In this Letter we show under what conditions, considering a nonlinear coupling
between system and a bath of TLSs, it is possible to derive a nonlinear a QLE for
the dynamics of the degrees of freedom of the system, having in mind a circuit
QED setup. In addition, we show how the nonlinear QLEs derived here can
represent an explanation to some of the phenomena recently observed in the
context of microwave quantum optomechanics
\cite{Pirkkalainen:2015ki}. 

The starting point for our analysis is represented by a bosonic system
$(\mathcal{S})$ coupled to an environment $(\mathcal{E})$ . The total
Hamiltonian of the bipartite system $(\mathcal{S}+\mathcal{E})$ is given by
\begin{equation}
 H=H_{\mathcal{S}}+H_{\mathcal{E}}+H_{\mathcal{S}-\mathcal{E}},
\label{eq:1}
\end{equation}
where $H_{\mathcal{S}}=H_{\mathcal{S}}(c,c^\dagger)$ is the Hamiltonian of the
isolated system, exhibiting a generic dependence on the annihilation (creation)
operators $c$ $(c^\dagger)$ associated with the system, and $H_{\mathcal{E}}$ is the
Hamiltonian for the bath. 

We assume here that the environment Hamiltonian can be decomposed into two
terms,
$H_\mathcal{E}^\mathrm{B}= \sum_\mathrm{k} \omega_\mathrm{k}
b^\dagger_\mathrm{k} b_\mathrm{k}$ and $H_\mathcal{E}^\mathrm{TLS}$,
corresponding to a bath of free bosonic modes, and to a bath of TLSs,
respectively. The bosonic bath describes, for instance, the modes of the
electromagnetic field of the environment. In our analysis we assume that these
modes, while being associated with the noise properties and dissipation of the
system, encompass also the external coherent fields driving the system whose
properties are encoded in the state of the bath for the modes
$b_\mathrm{k}$. Our choice is equivalent to considering a coherent driving term
for the system Hamiltonian and a purely thermal bath.

In this scenario, we describe the coupling between these modes and the degrees
of freedom of the system by the following Hamiltonian
\begin{align}
   H_\mathcal{\mathcal{S}-\mathrm{B}}= \sum g_\mathrm{k}^\mathrm{B}
  \left(c^\dagger b_\mathrm{k}+c b^\dagger_\mathrm{k}\right).
  \label{eq:7}
\end{align}
In addition, we model the bath of TLSs as a collection of spins
$\mathbf{J}_\mathrm{k}$. In this scenario we have that
$ H_\mathcal{E}^\mathrm{TLS}=\sum \Omega_\mathrm{k} J_{z}^{\mathrm{k}}$.  This
choice for the modeling of TLSs corresponds to the idea that, for each
$\Omega_\mathrm{k}$ multiple TLSs are present that collectively couple with the
system $\mathcal{S}$. While for $\Omega_\mathrm{k} \simeq \omega_\mathrm{S}$
--where $\omega_\mathrm{S}$ corresponds to a characteristic frequency for the
system-- the presence of impurities leads to a renormalization of the linewidth
associated with the linear response of the system induced by the
coupling given in Eq. \eqref{eq:7} (see Appendix \ref{sec:Fluct}); for
$\Omega_\mathrm{k} \simeq n \, \omega_\mathrm{S}$, nonlinear contributions
appear.  In our analysis, also in light of the recent investigations concerning
the relevance of two-photon emission processes by TLSs
\cite{Lindkvist:2014dd,Fischer:2017ca} when coupled to bosonic modes, we
consider the case $n=2$, representing the lowest-order approximation beyond
linear coupling.
\begin{figure}[htb]
  \centering
  \includegraphics[width=\columnwidth]{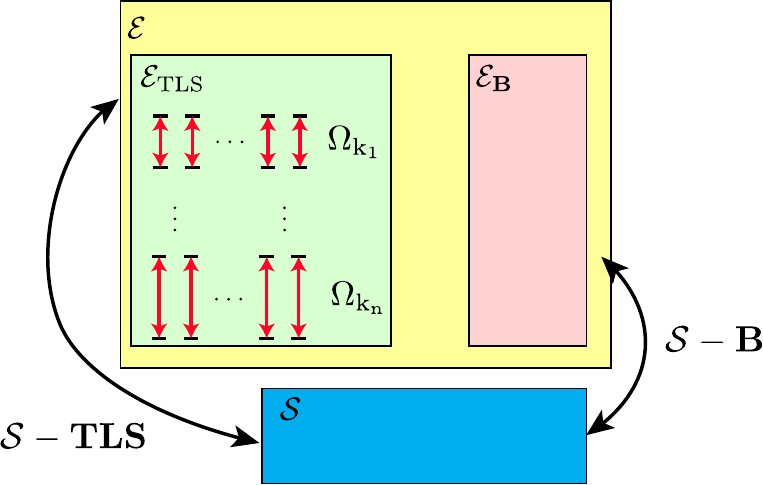}
  \caption{Cartoon picture of the setup. The system $\mathcal{S}$ is coupled to
    an environment $\mathcal{E}$, which is constituted by a bosonic bath
    $\mathcal{E}_\mathrm{B}$ and a bath of TLSs $\mathcal{E}_\mathrm{TLS}$. The
    coupling between the two baths and the system is mediated by the
    Hamiltonians $H_{\mathcal{S}-\mathrm{B}}$ and $H_{\mathcal{S}-\mathrm{TLS}}$,respectively. }     
  \label{fig:open_system}
\end{figure}
This assumption appears to be compatible with the usual experimental conditions
encountered in the context of circuit QED where microwave cavities operate at
frequencies corresponding to few GHz \cite{Wallraff:2004dy,Sillanpaa:2007ig,Grabovskij:2012bda}
while the energy separation of a TLS relevant for the physics of either of these
systems is of the order of $10\,$GHz \cite{Grabovskij:2012bda,Holder:2013gu}.
In this case, it is possible to write the system-TLS coupling
Hamiltonian as
\begin{align}
  H_{\mathcal{S}-\mathrm{TLS}}&= \sum_\mathrm{k} g^\mathrm{TLS}_\mathrm{k} \left(J^\mathrm{k}_+ c^2+J^\mathrm{k}_- {c^\dagger}^2\right). 
  \label{eq:8}
\end{align}

If we assume that $\left|\mathbf{J}_\mathrm{k}\right|\gg 1$, corresponding to
the idea that, for each value of $\mathrm{k}$ multiple TLSs couple to the system
$\mathcal{S}$, by resorting to the Holstein-Primakoff (HP) realization of spin
operators in terms of bosonic modes, we can replace the spin operators with
bosonic ones.  This mapping can be performed in two different ways,
corresponding to complementary experimental conditions
(see Appendix \ref{sec:HP}) . If it is assumed that the TLSs mostly reside in their
ground state, we have that $J^\mathrm{k}_3 \simeq -j_\mathrm{k}$ -- where
$j_\mathrm{k}$ is the index of the representation associated with the spin
$\mathbf{J}_\mathrm{k}$-- and the HP mapping reads
$J^\mathrm{k}_3 \to d^\dagger_\mathrm{k}d_\mathrm{k}-j_\mathrm{k}$,
$J^\mathrm{k}_-\to d_\mathrm{k}$, $J^\mathrm{k}_+\to d_\mathrm{k}^\dagger$.  In
this case, the coupling between the system and the TLS bath can be approximated
by
\begin{align}
  H_{\mathcal{S}-\mathrm{HP}_{-}}&= 
  \sum_\mathrm{k} g^\mathrm{HP}_\mathrm{k} 
  \left(d_\mathrm{k}^\dagger c^2+ d_\mathrm{k} {c^\dagger}^2\right). 
  \label{eq:10}
\end{align}
with $g^\mathrm{HP}_\mathrm{k}=\sqrt{2 j_\mathrm{k}} g^\mathrm{TLS}_\mathrm{k}$
On the other hand, if the TLSs mainly reside in their excited state
($J^\mathrm{k}_3 \simeq +j$) the mapping can be written as
$J^\mathrm{k}_3 \to j_\mathrm{k}-d^\dagger_\mathrm{k}d_\mathrm{k}$,
$J^\mathrm{k}_-\to d^\dagger_\mathrm{k}$, $J^\mathrm{k}_+\to d_\mathrm{k}$,
leading to the following approximation for $H_{\mathcal{S}-\mathrm{TLS}}$
\begin{align}
  H_{\mathcal{S}-\mathrm{HP}_{+}}&= \sum_\mathrm{k} 
  g^\mathrm{TLS}_\mathrm{k} \left(d_\mathrm{k} c^2+ d_\mathrm{k}^\dagger {c^\dagger}^2\right). 
  \label{eq:15}
\end{align}
These two different forms of the HP mapping correspond to two different physical
situations: in the former case, the TLSs prevalently reside in their ground
state corresponding to the idea that the impurities mainly reside in their
ground state, implying a low-temperature regime. In this case, the bosonic
excitations described by the operators $d_\mathrm{k}$, represent (weak)
excitations around the ground state. On the other hand, the latter case
corresponds to the situation in which the highest-excited (metastable) state of
the TLSs is weakly (de-)excited: corresponding, for instance, to the case in
which an external drive induces excitations in the TLSs bath, leading to a
possible interpretation of the linewidth narrowing observed in circuit QED
setups under strong driving conditions \cite{Martinis:2005kwa} in terms of
nonlinear QLEs associated with the saturation of the TLSs. In this picture, 
the external drive effectively heats the impurities to their excited state, inducing
the population inversion for the ensemble of TLSs and a consequent saturation,
justifying the $HP_{+}$ transformation in terms of (weak) de-excitations of the
highest-excited state.

As we show in Appendix \ref{sec:QLE-F}, it is possible to derive QLEs for the
system, provided that the environment Hamiltonian is described by a set of
bosonic operators, coupled linearly to the system degrees of freedom. It is
important to note that the requirement of linearity concerning the
system/environment Hamiltonian is limited to the bath degrees of freedom,
meaning that its most general form can be expressed as
\begin{align}
H_{\mathcal{ S}-\mathcal{E}}=\sum_{\mathrm{k}} g_{\mathrm{k}} 
\left[F^\dagger\left(c,c^\dagger\right) e_{\mathrm{k}}+
 F\left(c,c^\dagger\right) e_{\mathrm{k}}^\dagger \right],
  \label{eq:16}
\end{align}
where $e_\mathrm{k}$ and $e^\dagger_\mathrm{k}$ represent generic bosonic
operators associated with the environment degrees of freedom. The form the
system/environment coupling represents a sufficient condition for the derivation
of a nonlinear QLE, along with the assumption that the modes of the bath are
noninteracting. In other terms, it is necessary to assume a linear dependence of
the coupling Hamiltonian on the environment degrees of freedom since, in order
to derive the QLEs for the system, the solution of the Heisenberg equation of
motion for the environment degrees of freedom has to assume a specific form, in
which the contribution of the system and the environment operators can be
represented as two separate additive terms (see Appendix \ref{sec:QLE-F}) .

It is therefore clear that, since the form of $H_{\mathcal{S}-\mathrm{B}}$ and
$H_{\mathcal{S}-\mathrm{HP}_\pm}$ can be expressed in the form given by
Eq. \eqref{eq:16}, with $F\left(c,c^\dagger\right)$ given by $c$, $c^2$ and
${c^\dagger}^2$, and with
$e_\mathrm{k}=b_\mathrm{k}$ and $e_\mathrm{k}=d_\mathrm{k}$ for $\mathcal{S}-\mathrm{B}$,
$\mathcal{S}-\mathrm{HP}_-$,$\mathcal{S}-\mathrm{HP}_+$ respectively, we can write the
dynamics of the system in terms of a (nonlinear) QLE as
\begin{subequations}
  \begin{align}
    \dot{c} =-i \left[c, H_\mathcal{S} \right] &-\left(\frac{\kappa}{2} + \kappa_N
    c^\dagger c \right) c  \nonumber \\ &+\sqrt{\kappa} c_\mathrm{in} + 2 \sqrt{\kappa_N}
    c^\dagger c^{\mathrm{TLS}}_\mathrm{in}  
    \label{eq:11} \\
    \dot{c} =-i \left[c, H_\mathcal{S} \right] &-\left(\frac{\kappa}{2} - \kappa_N
    c^\dagger c \right) c  \nonumber \\ &+\sqrt{\kappa} c_\mathrm{in} + 2 \sqrt{\kappa_N}
    c^\dagger {c_\mathrm{in}^{\mathrm{TLS}}}^\dagger.
    \label{eq:12}
  \end{align}
\end{subequations}
Eqs.~(\ref{eq:11},\ref{eq:12}), obtained considering the system/environment
coupling given by $H_{\mathcal{S}-\mathrm{HP}_{-}}$ and
$H_{\mathcal{S}-\mathrm{HP}_{+}}$ respectively, are the main result of our
analysis: the presence of a TLS bath leads to the appearance of nonlinear
dissipative terms ($\pm \kappa_\mathrm{N} c^\dagger c\,c$), and to purely imaginary
parametric noise terms ($2\sqrt{\kappa_{\rm N}} c^\dagger {c_{\rm in}^{\rm
    TLS}}^{(\dagger)}$). We stress here that these terms are the direct result of the
modelling of the bath in terms of two-separate environments
($H_{\mathcal{S}-\mathrm{B}}$ and $H_\mathcal{S}-\mathrm{HP_\pm}$), and do not
represent an ad-hoc modification of the linear QLEs that can be derived in the
absence of  coupling to TLSs. In particular, while the nonlinear dissipation
term possibly represents a natural extension to the nonlinear regime of linear
QLEs, the parametric noise term is a nontrivial contribution associated
with the presence of the TLS bath.

In addition we observe here that, analogously to their linear counterpart,
Eqs.~\eqref{eq:12} are time-local, i.e. the dynamics is Markovian. As detailed
in Appendix \ref{sec:QLE-F}, this property is related to the assumption that,
within the range of frequencies of interest, the coupling strength between
system and environment is independent of the mode considered (wide band limit
approximation) \cite{Stefanucci:2014uu}.  
  
\begin{figure}[htb]
  \centering
  \includegraphics[width=\columnwidth]{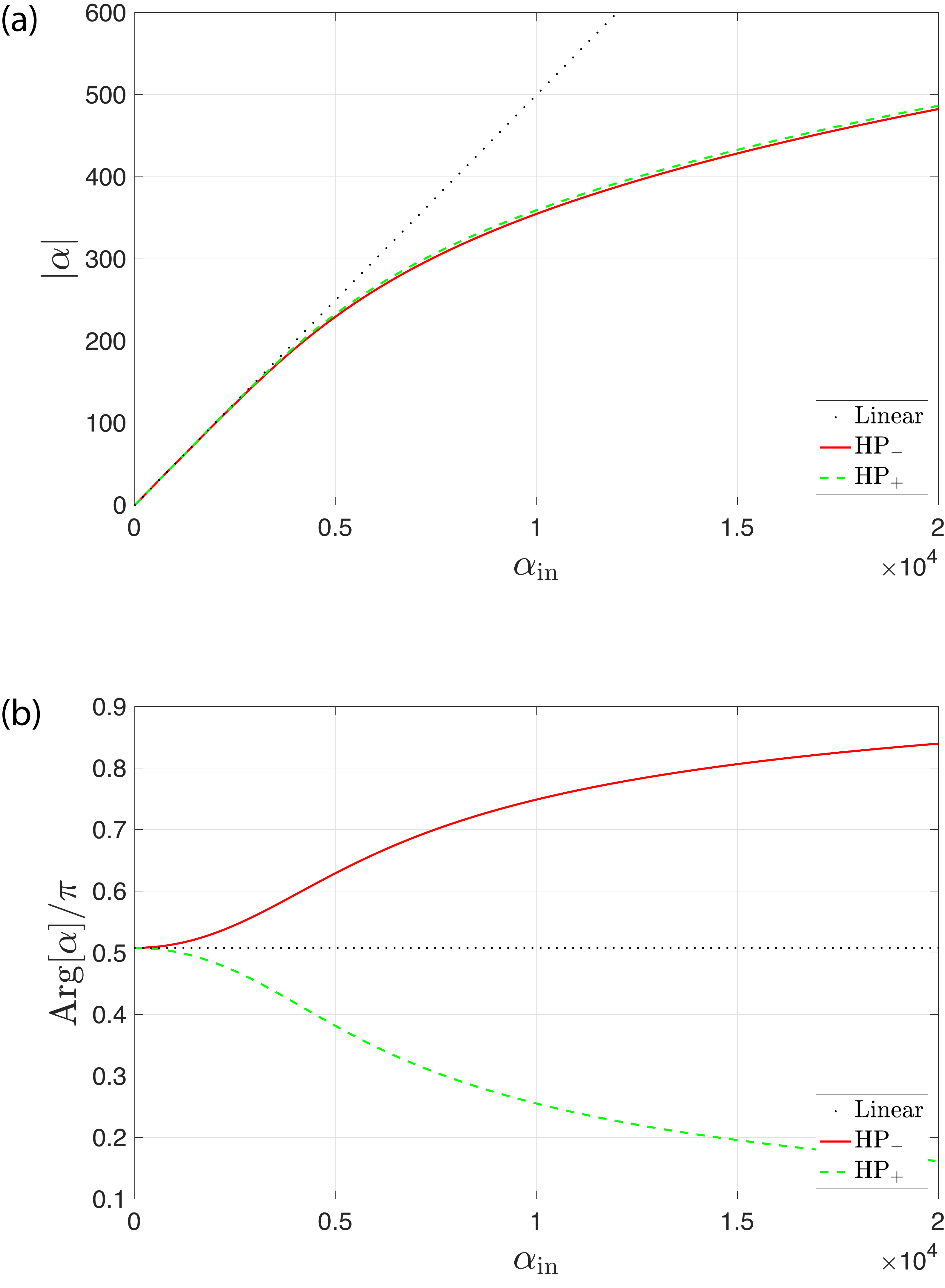}
  \caption{Amplitude (a) and phase (b), for the stationary value (in a frame
    rotating at $\omega_\mathrm{p}$, see text) of the cavity
    field $\alpha$ in the presence of a driving
    $\alpha_\mathrm{in}$. Parameters: $\kappa_\mathrm{N}=1.5 \times 10^{-4}$,
    $\Delta=20$ (all quantities are expressed in units of $\kappa$).} 
  \label{fig:st_sol}
\end{figure}

If we further consider a pump-probe representative of a circuit QED setup
(e.g. a circuit optomechanical experiment), we can assume that the dynamics
given by Eq. \eqref{eq:14} is linearized around a strong coherent tone
$$
\alpha_\mathrm{p}=\alpha_\mathrm{in} \exp\left[-i \omega_\mathrm{p} t\right].
$$
The frequency $\omega_\mathrm{p}$ is detuned by
$\Delta=\omega_\mathrm{p}-\omega_\mathrm{c}$ from the cavity resonant frequency.
As a result of the linearization scheme, we have that the amplitude of the
cavity field oscillating at $\omega_\mathrm{p}$ is given by the solution of a
nonlinear algebraic equation. In Fig. \ref{fig:st_sol} we have plotted the
stationary value of the cavity field for the two choices of the HP mapping
($\mathrm{HP}_\pm$). As expected, for small values of the driving field $\alpha_\mathrm{in}$,
the stationary solution corresponds to the solution in the absence of nonlinear
dissipation. However, for larger values of $\alpha_\mathrm{in}$ the stationary
solution substantially deviates from the solution of the linear system, with,
for the parameters discussed here, a negligible difference between
$\mathrm{HP}_\pm$ cases.

Furthermore, the (first-order) dynamics of the fluctuations $c = \alpha + a $
around the stationary value induced by the pump (in a frame rotating at
$\omega_\mathrm{p}$) is given by
\begin{subequations}
  \begin{align}
    \dot{a}=&\left[i \Delta -\left(\frac{\kappa}{2}  + 2 \kappa_N \left|\alpha\right|^2\right)\right] a - \kappa_N \alpha^2 a^\dagger
              \nonumber \\
            &    +  \sqrt{\kappa} a_\mathrm{in} +2 \sqrt{\kappa_N}\alpha^*
              a^{\mathrm{TLS}}_\mathrm{in} \label{eq:13} \\
              \dot{a}=&\left[i \Delta -\left(\frac{\kappa}{2}  - 2 \kappa_N \left|\alpha\right|^2\right)\right] a + \kappa_N \alpha^2 a^\dagger
                        \nonumber \\
            &    +  \sqrt{\kappa} a_\mathrm{in} +2 \sqrt{\kappa_N}\alpha^*
               {a_\mathrm{in}^{\mathrm{TLS}}}^\dagger
              \label{eq:14}
  \end{align}
\end{subequations}
$\mathrm{HP}_-$ and $\mathrm{HP}_+$ case respectively
(see Appendix \ref{sec:Lin}). It is possible to see that
Eqs.~(\ref{eq:13},\ref{eq:14}) include a purely imaginary parametric term, on
top of a nonlinear dissipation term implying linewidth broadening or narrowing,
depending on the state of the TLSs bath. Recently, in Ref. \cite{Pirkkalainen:2015ki} a
term of the same form was introduced as an ad-hoc parameter, in order 
to match the experimental results of a cavity optomechanical experiment aimed at
establishing squeezing below the SQL of a nanomechanical
resonator.

\begin{figure}[htb]
  \centering
  \includegraphics[width=\columnwidth]{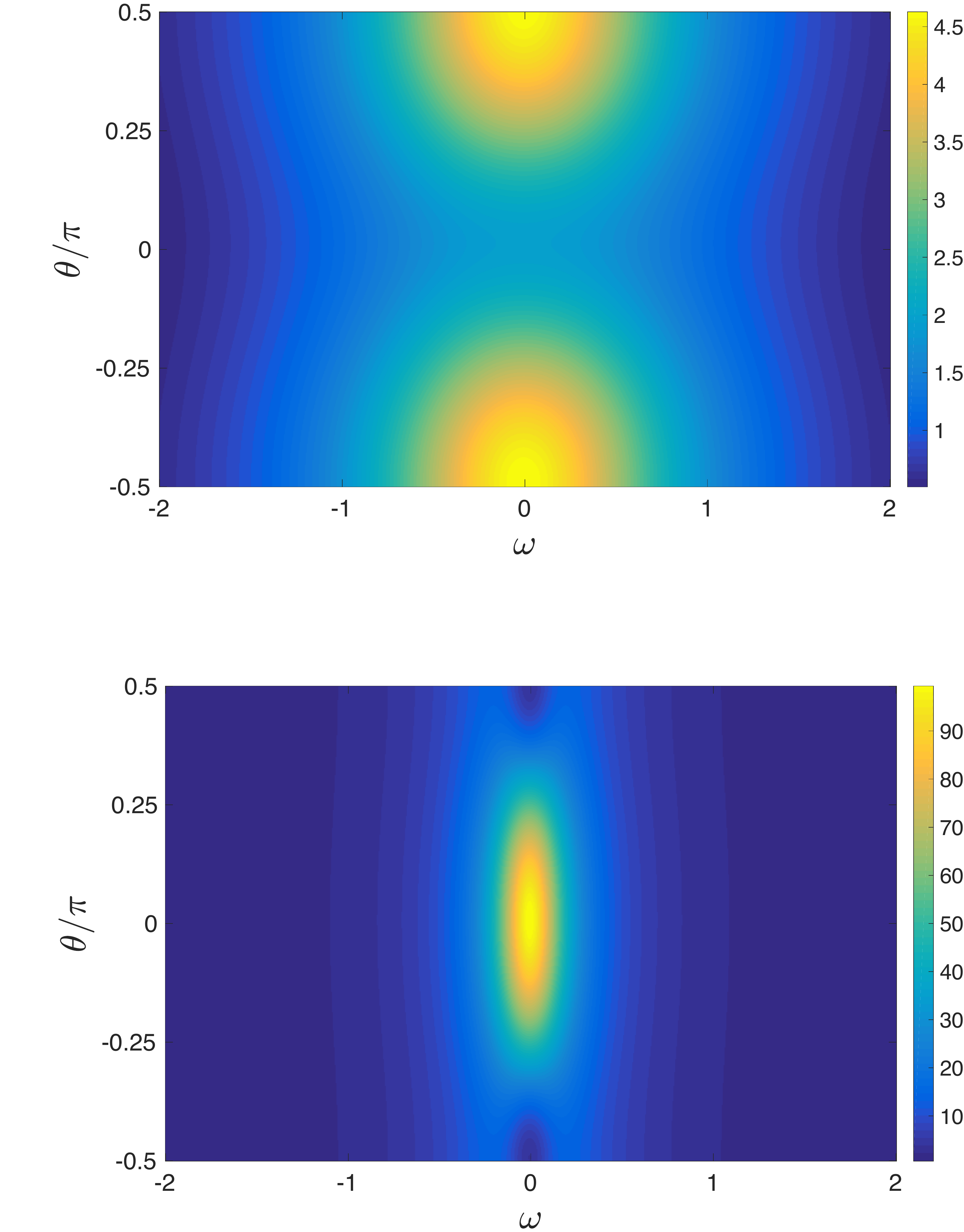}
  \caption{Noise spectrum for the cavity field in the presence of an external
    drive $\alpha_\mathrm{in}=700$, for (a) $\mathrm{HP}_-$ and (b)
    $\mathrm{HP}_+$ for $\langle a^\dagger_\mathrm{in}
    a_\mathrm{in}\rangle=
    \langle {a_\mathrm{in}^\mathrm{TLS}}^\dagger
    a^\mathrm{TLS}_\mathrm{in}\rangle=1$ (all other parameters as in Fig. \ref{fig:st_sol}).}
  \label{fig:noise}
\end{figure}

Our description, therefore provides a potential explanation of such parametric
effects in terms of nonlinear dissipation phenomena associated with the
nonlinear coupling to a bath of TLSs.  In order to characterize the effect
induced by the presence of the nonlinear coupling to TLSs, we evaluate the
fluctuation spectrum of the cavity field
$S_\omega^{\theta}=1/2 \langle\left\{X^\theta_\omega,
  X^\theta_{-\omega}\right\}\rangle$ -- with
$X_\omega^\theta = 1/\sqrt{2} \left(a_{-\omega}^\dagger e^{i \theta}+a_\omega
  e^{-i \theta}\right)$ -- assuming thermal fluctuations both for the bosonic
and the TLS bath. As hinted by the structure of Eqs.~(\ref{eq:13},\ref{eq:14}),
the presence of a parametric term induces squeezing, which can be experimentally
observed by homodyne detection of the output field, in the cavity spectrum for
both cases, as it is possible to see from Fig. \ref{fig:noise} where it is
possible to see how the cavity fluctuation spectrum exhibits a clear dependence
on the phase $\theta$.

We have reported here how it is possible to deduce nonlinear QLEs for the
dynamics of an open quantum system from a nonlinear system/environment coupling
Hamiltonian. Moreover, we have discussed how an effective nonlinear
system/environment coupling can emerge in the presence of impurities modeled as
TLSs. Ultimately, we have shown that the TLS-induced nonlinearities can
represent a potential explanation for the imaginary parametric terms reported in
Ref. \cite{Pirkkalainen:2015ki}.

This work was supported by the Academy of Finland (Contract No. 275245). 
\onecolumngrid 
\appendix 

\section{Holstein-Primakoff transformation}
\label{sec:HP}

We discuss here the Holstein-Primakoff realization allowing us to replace the
spin operators $J_z$, $J_\pm$ obeying the usual SU(2) commutation relations
\begin{equation}
\label{eq:S31}
	\left[J_z^\mathrm{k},J_\pm^\mathrm{k}\right] = \pm J_\pm^\mathrm{k}, \qquad \left[J_+^\mathrm{k},J_-^\mathrm{k}\right] = 2 J_z^\mathrm{k},
\end{equation} 
with bosonic operators $d_\mathrm{k}$, $d_\mathrm{k}^\dagger$, for which
\begin{equation}
\label{eq:S36}
	\left[d_\mathrm{k}, d_\mathrm{k}^\dagger\right] = 1 .
\end{equation}
As discussed in the main text, in order to map the spin operators obeying
Eq. \eqref{eq:S31} with the bosonic operators $d_\mathrm{k}$,
$d_\mathrm{k}^\dagger$, we have two possibilities, depending on the physical
situation we want to describe. If we assume that $J_z^\mathrm{k} \simeq -j_{\rm k}$
--this choice is indicated in the main text as $\mathrm{HP}_-$--, we
can consider the following transformation
\begin{equation}
\label{eq:S30}
	J_z^\mathrm{k} = \hat{n}_{\rm k} - j_\mathrm{k} , \qquad
	J_+^\mathrm{k} = d_\mathrm{k}^\dagger \sqrt{2j_\mathrm{k} - \hat{n}_{\rm k}} , \qquad
	J_-^\mathrm{k} = \sqrt{2j_\mathrm{k} - \hat{n}_{\rm k}}\, d_\mathrm{k} ,
\end{equation} 
where $\hat{n}_{\rm k} = d_\mathrm{k}^\dagger d_\mathrm{k}$. The operators
$J_z^\mathrm{k}$, $J_\pm^\mathrm{k}$ can be shown to fulfill the SU(2)
commutation relations
\begin{subequations}
\label{eq:S37}
\begin{align}
	&\left[J_z^\mathrm{k}, J_+^\mathrm{k}\right] = \left[\hat{n}_{\rm k},d_\mathrm{k}^\dagger\right] \sqrt{2j_\mathrm{k} - \hat{n}_{\rm k}} = J_+^\mathrm{k} , \qquad
	\left[J_z^\mathrm{k}, J_-^\mathrm{k}\right] =  \sqrt{2j_\mathrm{k} - \hat{n}_{\rm k}} \left[\hat{n}_{\rm k},d_\mathrm{k}\right] = -J_-^\mathrm{k} ,\\
	&\left[J_+^\mathrm{k}, J_-^\mathrm{k}\right] = d_\mathrm{k}^\dagger \left(\sqrt{2j_\mathrm{k} - \hat{n}_{\rm k}}\right)^2 d_\mathrm{k} - \sqrt{2j_\mathrm{k} - \hat{n}_{\rm k}} \,\hat{n}_{\rm k}\, \sqrt{2j_\mathrm{k} - \hat{n}_{\rm k}} = \hat{n}_{\rm k} \left(2j_\mathrm{k} -\hat{n}_{\rm k}+1\right) - 2j_\mathrm{k} + \hat{n}_{\rm k} - \hat{n}_{\rm k} \left(2j_\mathrm{k}-\hat{n}_{\rm k}\right) = 2 J_z^\mathrm{k} .
\end{align}
\end{subequations}
In the limit $j_\mathrm{k} \to \infty$, we have that
\begin{equation}
\label{eq:S38}
	\frac{J_+^\mathrm{k}}{\sqrt{2j}} = d_\mathrm{k}^\dagger \sqrt{\frac{2j- \hat{n}_{\rm k}}{2j_\mathrm{k}}} = d_\mathrm{k}^\dagger \left(1 - \frac{\hat{n}_{\rm k}}{4j_\mathrm{k}} + \ldots\right) \simeq d_\mathrm{k}^\dagger, \qquad \frac{J_-^\mathrm{k}}{\sqrt{2j}} \simeq d_\mathrm{k} , \qquad \frac{J_z^\mathrm{k}}{j} = \frac{\hat{n}_{\rm k}}{j_\mathrm{k}} - 1 \simeq -1 .
\end{equation}
Therefore the bosonic excitations described by $d_\mathrm{k}$ and
$d_\mathrm{k}^\dagger$ correspond to (small) excitations around the
$J_z^\mathrm{k} = -j_\mathrm{k}$ state. Conversely we can write
\begin{equation}
\label{eq:S34}
	J_z^\mathrm{k} = j_\mathrm{k} - \hat{n}_{\rm k} ,\qquad
	J_-^\mathrm{k} = d_\mathrm{k}^\dagger \sqrt{2j_\mathrm{k}-\hat{n}_{\rm k}} ,\qquad
	J_+^\mathrm{k} = \sqrt{2j_\mathrm{k}-\hat{n}_{\rm k}}\, d_\mathrm{k} .
\end{equation} 
so that when $j_\mathrm{k} \to \infty$
\begin{equation}
\label{eq:S38}
\frac{J_+^\mathrm{k}}{\sqrt{2 j_\mathrm{k}}}\simeq d_\mathrm{k}, \qquad \sqrt{\frac{2}{j_\mathrm{k}}}J_-^\mathrm{k} \simeq d_\mathrm{k}^\dagger , \qquad \frac{J_z^\mathrm{k}}{j_\mathrm{k}} = 1 - \frac{\hat{n}_{\rm k}}{j_\mathrm{k}} \simeq 1 ,
\end{equation}
which correspond to the description of small fluctuations around the
$J_z^\mathrm{k} = j$ state, indicated as $\mathrm{HP}_+$ in main text.

 
\section{QLE for $F(c,c^\dagger)$}
\label{sec:QLE-F}
We discuss here the form of the QLEs generated by a model for which, following
the notation introduced in Eq. (1) of the main text, $H_\mathcal{S}$ is left
unspecified, the environment is given by a set of noninteracting bosonic modes
described by
$H_\mathcal{E} = \sum_\mathrm{k} \omega_\mathrm{k} e_\mathrm{k}^\dagger
e_\mathrm{k}$, where $e_\mathrm{k}$ $(e_\mathrm{k}^\dagger)$ are the
annihilation (creation) operators associated with mode k and the
system/environment coupling is given by the following Hamiltonian
\begin{equation}
\label{eq:S16}
	H_{\mathcal{S}-\mathcal{E}} = \sum_\mathrm{k} g_\mathrm{k} \left[F(c,c^\dagger) e^\dagger_\mathrm{k} + F^\dagger(c,c^\dagger) e_\mathrm{k}\right] ,
\end{equation}
where $F(c,c^\dagger)$ is a generic function of the creation and annihilation operators of
the system. Since $H_{\mathcal{S}-\mathcal{E}}$ is a linear operator with
respect to the degrees of freedom of the bath, and $e_\mathrm{k}^{(\dagger)}$
commutes with $H_\mathcal{S}$, we can follow the same strategy employed for the
derivation of the linear QLEs \cite{Wiseman:2010vw} and write the equations of
motion (EOM) for the
bath field operators in the Heisenberg picture as
\begin{equation}
\label{eq:S17}
	\dot{e}_\mathrm{k}(t)=-i\omega_\mathrm{k} e_\mathrm{k}(t) - i g_\mathrm{k} F(c,c^\dagger) .
\end{equation}
Similarly, the EOM for the system can be written as
\begin{equation}
\label{eq:S18}
	\dot{c}(t)=i[H_\mathcal{S},c(t)] + i\sum_\mathrm{k} g_\mathrm{k} \left([F,c]e^\dagger_\mathrm{k} + [F^\dagger,c]e_\mathrm{k}\right) .
\end{equation}
Equation \eqref{eq:S17} can be solved in terms of an initial condition $t_0$, yielding
\begin{equation}
\label{eq:S19}
	e_\mathrm{k}(t) = e^{-i\omega_\mathrm{k}(t-t_0)} e_\mathrm{k}(t_0) - i g_\mathrm{k} \int_{t_0}^t e^{-i\omega_\mathrm{k}(t-t')} F\left(c(t'),c^\dagger(t')\right)\,\mathrm{d}t' .
\end{equation}
By substituting Eq. \eqref{eq:S19} and its Hermitian conjugate into
Eq. \eqref{eq:S18} we obtain
\begin{equation}
\label{eq:S21}
\begin{split}
  \dot{c}(t) = i[H_\mathcal{S},c(t)] + i &\sum_\mathrm{k} g_\mathrm{k}\bigg\lbrace [F,c]\bigg[e^{i\omega_\mathrm{k}(t-t_0)}e_\mathrm{k}^\dagger(t_0) + ig_\mathrm{k}\int_{t_0}^t e^{i\omega_\mathrm{k}(t-t')}F^\dagger(t')\,\mathrm{d}t'\bigg]\\
  &+[F^\dagger,c]\bigg[e^{-i\omega_\mathrm{k}(t-t_0)}e_\mathrm{k}(t_0) -i
  g_\mathrm{k}\int_{t_0}^t
  e^{-i\omega_\mathrm{k}(t-t')}F(t')\,\mathrm{d}t'\bigg]\bigg\rbrace .
\end{split}
\end{equation}
Like for the purely linear case, we introduce the density of states
$D = \partial \mathrm{k}/\partial \omega_\mathrm{k}$ (supposing a continuum of
states for the bath) and assume that, in the relevant frequency regime,
$g_\mathrm{k}$ does not depend on the mode index k. If we define
\begin{equation}
\label{eq:S22}
g_\mathrm{k}=\sqrt{\frac{\kappa }{2\pi D}} ,
\end{equation}
where $\kappa$ a the mode-independent  constant, we can write Eq. \eqref{eq:S21} as
\begin{equation}
\label{eq:S23}
\begin{split}
	\dot{c}(t) &= i[H_\mathcal{S},c(t)] + i \sum_\mathrm{k} \sqrt{\frac{\kappa }{2\pi D}} \bigg\lbrace [F,c]\bigg(e^{i\omega_\mathrm{k}(t-t_0)}e_\mathrm{k}^\dagger(t_0) + i\sqrt{\frac{\kappa }{2\pi D}} \int_{t_0}^t e^{i\omega_\mathrm{k}(t-t')}F^\dagger(t')\,\mathrm{d}t'\bigg)\\
	&\qquad \qquad \qquad \qquad+ [F^\dagger,c]\bigg(e^{-i\omega_\mathrm{k}(t-t_0)}e_\mathrm{k}(t_0) - i \sqrt{\frac{\kappa }{2\pi D}}\int_{t_0}^t e^{-i\omega_\mathrm{k}(t-t')}F(t')\,\mathrm{d}t'\bigg)\bigg\rbrace\\
	&= i[H_\mathcal{S},c(t)] +\sqrt{\kappa} \bigg\lbrace [F,c]\bigg(-c_\mathrm{in}^\dagger(t)- \frac{\sqrt{\kappa}}{2} F^\dagger(t)\bigg) + [F^\dagger,c]\bigg(-c_\mathrm{in}(t) +  \frac{\sqrt{\kappa}}{2}F(t)\bigg)\bigg\rbrace ,
\end{split}
\end{equation}
where we have defined $c_\mathrm{in}(t)$  as
\begin{equation}
\label{eq:S24}
c_{\mathrm{in}} (t)= -\frac{i}{\sqrt{2 \pi D}} \sum_k e^{-i \omega_k \left( t-t_0 \right)} e_k \left( t_0 \right) .
\end{equation}
The definition introduced in Eq.~\eqref{eq:S22} corresponds to what, in the
context of electronic transport is defined as ``wide band limit approximation''
and, allowing us to write the QLE given in Eq.~\eqref{eq:S23} in time-local
form, can be considered equivalent to the Markov approximation \cite{Stefanucci:2014uu}.
  
Let us focus on the case, discussed in the text, of two separate baths:  a bosonic bath with operators
$b_\mathrm{k}$ and a bath of TLSs with HP-transformed modes $d_\mathrm{k}$. We
define two functions $F_\mathrm{b}$ and $F_\mathrm{TLS}$ of the system operators
that couple to the bosonic and TLS baths, respectively. The QLE \eqref{eq:S23} then
reads
\begin{equation}
\label{eq:S25}
\begin{split}
	\dot{c}(t) = i[H_\mathcal{S},c(t)] &+\sqrt{\kappa} \bigg\lbrace [F_\mathrm{b},c]\bigg(-c_\mathrm{in}^\dagger- \frac{\sqrt{\kappa}}{2} F_\mathrm{b}^\dagger\bigg) + [F_\mathrm{b}^\dagger,c]\bigg(-c_\mathrm{in} +  \frac{\sqrt{\kappa}}{2}F_\mathrm{b}\bigg)\bigg\rbrace \\
	&+\sqrt{\kappa_\mathrm{N}} \bigg\lbrace [F_\mathrm{TLS},c]\bigg(-c_\mathrm{in}^{\mathrm{TLS}\dagger} - \frac{\sqrt{\kappa_\mathrm{N}}}{2} F_\mathrm{TLS}^\dagger\bigg) + [F_\mathrm{TLS}^\dagger,c]\bigg(-c_\mathrm{in}^\mathrm{TLS} +  \frac{\sqrt{\kappa_\mathrm{N}}}{2}F_\mathrm{TLS}\bigg)\bigg\rbrace .
\end{split}
\end{equation}
Assuming a linear coupling between the system and the bosonic bath and choosing
the $\mathrm{HP}_-$ mapping for the TLSs, one obtains $F_\mathrm{b} = c$ and
$F_\mathrm{TLS} = c^2$. Substituting these into Eq. \eqref{eq:S25} gives
\begin{equation}
\label{eq:S26}
	\dot{c} = i[H_\mathcal{S},c(t)] - \left(\frac{\kappa}{2} + \kappa_\mathrm{N}  c^\dagger c \right) c + \sqrt{\kappa} c_\mathrm{in} + 2 \sqrt{\kappa_\mathrm{N}} c^\dagger c_\mathrm{in}^\mathrm{TLS} .
\end{equation}
which corresponds to Eq. (7a) of the main text. On the
contrary, if the $\mathrm{HP}_+$ mapping is chosen, one
obtains Eq. (7b) with $F_\mathrm{TLS} =  c^{\dagger2}$.


\section{Linearization of the quantum Langevin equations}
\label{sec:Lin}
Here we outline the linearization strategy that allows us, in the presence of a
strong coherent tone
$\alpha_\mathrm{p} = \alpha_\mathrm{in} e^{-i \omega_\mathrm{p} t}$, to recast
Eqs. (7a,7b) of the main text in terms of equations describing the stationary
state (in a frame rotating at $\omega_\mathrm{p}$) and the fluctuations around
this stationary state, given by Eqs. (8a,8b) of the main text.

Focusing on Eq. (7a)
\begin{equation}
\label{eq:S1}
	\dot{c} = -i \left[c, H_\mathcal{S} \right] -\left(\frac{\kappa}{2} + \kappa_N
	c^\dagger c \right) c + \sqrt{\kappa} c_\mathrm{in} + 2 \sqrt{\kappa_N}
	c^\dagger c^\mathrm{TLS}_\mathrm{in}  
\end{equation}
in the presence of a strong coherent pump
$\alpha_\mathrm{p} = \alpha_\mathrm{in} e^{-i \omega_\mathrm{p} t}$, we seek a
solution of the form $c = \alpha + a$
\begin{equation}
\label{eq:S3}
-i \omega_\mathrm{p} \alpha + \dot{a} = -i \omega_\mathrm{c} \left(\alpha + a\right) - \left[\frac{\kappa}{2} + \kappa_N \left(\alpha^* + a^\dagger\right)\left(\alpha + a\right)\right] \left(\alpha + a\right) + \sqrt{\kappa} \left(\alpha_\mathrm{in} + a_\mathrm{in}\right) + 2\sqrt{\kappa_N} \left(\alpha^* + a^\dagger\right) a^\mathrm{TLS}_\mathrm{in} ,
\end{equation}
where without loss of generality, we have assumed that
$H_\mathcal{S} = \omega_\mathrm{c} c^\dagger c$.

Neglecting the fluctuation terms, we obtain the equation for the steady-state
solution
\begin{equation}
\label{eq:S4}
	0 = i \Delta \alpha - \frac{\kappa}{2} \alpha - \kappa_N \alpha \left|\alpha\right|^2 + \sqrt{\kappa} \alpha_\mathrm{in} ,
\end{equation}
where $\Delta = \omega_\mathrm{p} - \omega_\mathrm{c}$. From Eq. \eqref{eq:S3}
the equation for the fluctuation around the steady-state solution value of
$\alpha$ given above is thus expressed as
\begin{equation}
\label{eq:S5}
	 \dot{a} = \left[i \Delta -\left(\frac{\kappa}{2}  + 2 \kappa_N \left|\alpha\right|^2\right)\right] a - \kappa_N \alpha^2 a^\dagger + \sqrt{\kappa} a_\mathrm{in} + 2 \sqrt{\kappa_N}\alpha^* a^\mathrm{TLS}_\mathrm{in} .
\end{equation}
With a similar procedure one can also show that Eq. (7b) leads to Eq. (8b). Notice that the nonlinear dissipative terms
$\mp 2 \kappa_N \left|\alpha\right|^2 a$ in Eqs. (8a,8b) lead to the
broadening/narrowing of the linewidth associated with the linearized response of
the cavity field fluctuations, respectively.

\begin{figure}[tbh]
	\includegraphics[width=0.7\linewidth]{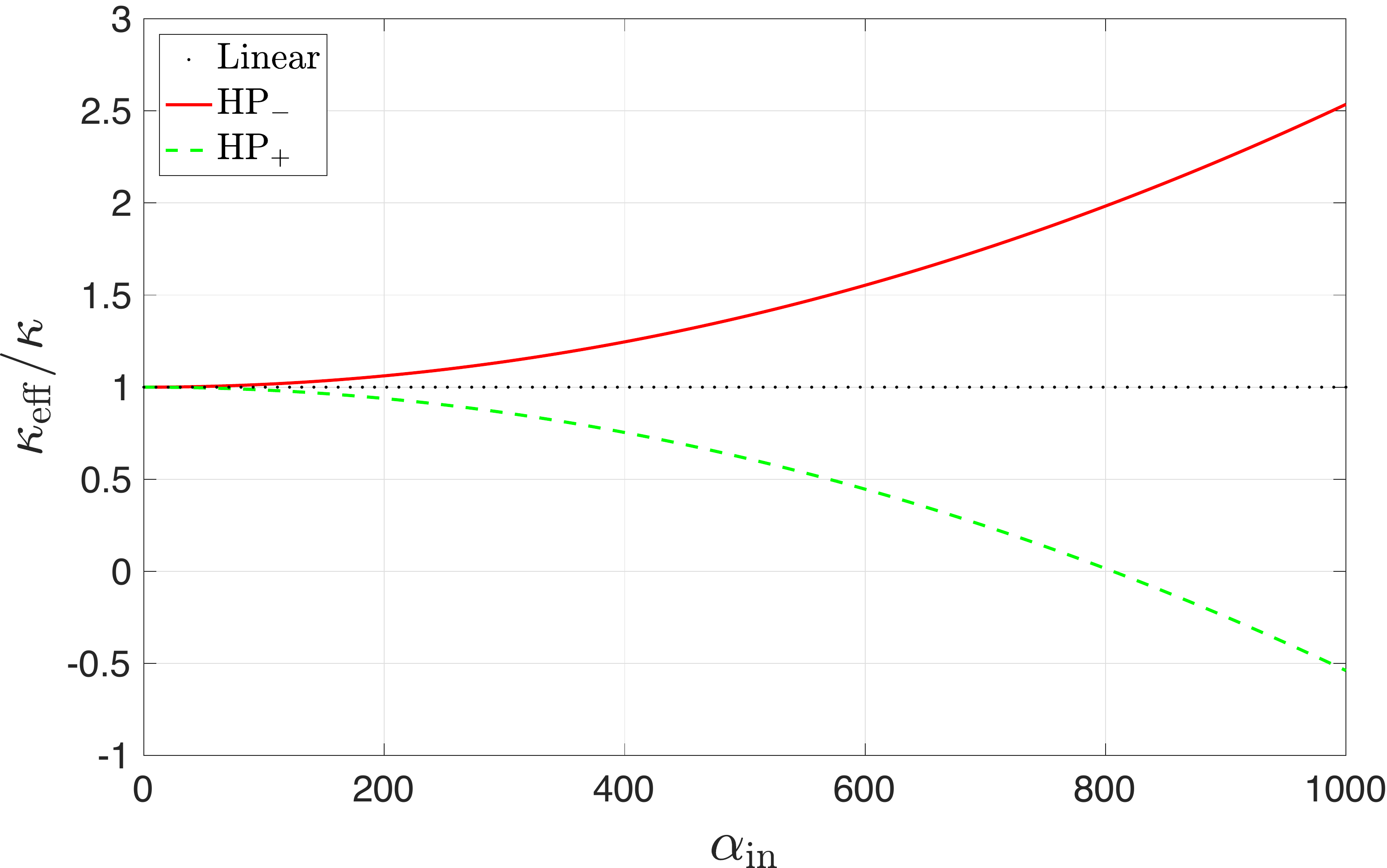}
	\caption{The total effective dissipation of the linearized models
          Eq. (8a) (red) and Eq. (8b) (dashed green) that correspond to the
          cases, where the majority of the TLSs are in the ground state/excited
          state, respectively. They are compared to the case of pure linear
          dissipation (black dots). Here we assume the system to be a simple
          cavity with $H_\mathcal{S} = \omega_\mathrm{c} c^\dagger c$. In the
          units of $\kappa$, the parameters are
          $\Delta = \omega_\mathrm{p} - \omega_\mathrm{c} = 20$ and
          $\kappa_N = 1.5 \times 10^{-4}$.}
	\label{fig:Keff}
\end{figure}


\section{Fluctuation spectrum of the nonlinear model}
\label{sec:Fluct}

Assuming that, in addition to the strong coherent tone, the dynamics of the
system is affected by thermal fluctuations of both the bosonic and the TLS baths
degrees of freedom, we evaluate here the spectrum of these fluctuations focusing
on the $\mathrm{HP}_-$ case (an analogous derivation holds for the
$\mathrm{HP}_+$ mapping). The fluctuation spectrum
\begin{equation}
\label{eq:S11}
S_\omega^\theta = \frac{1}{2} \Braket{\left\{X_\omega^\theta , X_{-\omega}^\theta \right\}} ,
\end{equation}
with
$X_\omega^\theta = 1/\sqrt{2} \left(a_{-\omega}^\dagger e^{i \theta} + a_\omega
  e^{-i \theta}\right)$, can be obtained by Fourier transforming the QLE given
by Eq. (8a) and its Hermitian conjugate
\begin{subequations}
\begin{align}
	&\left[-i \left(\omega + \Delta\right) + \frac{\kappa}{2} + 2\kappa_N \left|\alpha\right|^2 \right] a_\omega + \kappa_N \alpha^2 a_{-\omega}^\dagger = \sqrt{\kappa} a_{\mathrm{in},\omega} + 2 \sqrt{\kappa_N} \alpha^* a_{\mathrm{in},\omega}^\mathrm{TLS} \label{eq:S6} \\
	&\left[-i \left(\omega - \Delta\right) + \frac{\kappa}{2} + 2\kappa_N \left|\alpha\right|^2 \right] a_{-\omega}^\dagger + \kappa_N \alpha^{*2} a_{\omega} = \sqrt{\kappa} a_{\mathrm{in},-\omega}^\dagger + 2 \sqrt{\kappa_N} \alpha a_{\mathrm{in},-\omega}^{\mathrm{TLS}\dagger} \label{eq:S7}
\end{align}
\end{subequations}
with the usual convention for the Fourier transform, according to which
$a_t \xmapsto{\mathrm{FT}} a_\omega$ and
$a_t^\dagger \xmapsto{\mathrm{FT}} a^\dagger_{-\omega}$.

Defining
\begin{subequations}
\label{eq:S8a}
\begin{align}
	A &= -i \left(\omega + \Delta\right) + \frac{\kappa}{2} + 2\kappa_N \left|\alpha\right|^2 , \\
	B &= \kappa_N \alpha^2 , \\
	C &= -i \left(\omega - \Delta\right) + \frac{\kappa}{2} + 2\kappa_N \left|\alpha\right|^2 ,
\end{align}
\end{subequations}
the QLE for the system can be expressed as
\begin{equation}
\label{eq:S8b}
	\begin{pmatrix}
	a_\omega \\ a_{-\omega}^\dagger
	\end{pmatrix}
	= \frac{1}{AC - \left|B\right|^2}
	\begin{pmatrix}
	C & -B\\
	-B^* & A
	\end{pmatrix}
	\begin{pmatrix}
	\sqrt{\kappa} a_{\mathrm{in},\omega} + 2 \sqrt{\kappa_N} \alpha^* a_{\mathrm{in},\omega}^\mathrm{TLS}\\
	\sqrt{\kappa} a_{\mathrm{in},-\omega}^\dagger + 2 \sqrt{\kappa_N} \alpha a_{\mathrm{in},-\omega}^{\mathrm{TLS}\dagger} 
	\end{pmatrix} 
\end{equation}
and
\begin{subequations}
\begin{align}
	a_\omega &= \chi_\mathrm{d}\left(\omega\right) a_{\mathrm{in},\omega} + \chi_\mathrm{x}\left(\omega\right) a_{\mathrm{in},-\omega}^\dagger + \chi_\mathrm{d}^\mathrm{TLS} \left(\omega\right) a_{\mathrm{in},\omega}^\mathrm{TLS} + \chi_\mathrm{x}^\mathrm{TLS} \left(\omega\right) a_{\mathrm{in},-\omega}^{\mathrm{TLS}\dagger} , \label{eq:S9}\\
	a_{-\omega}^\dagger &= \chi_\mathrm{x}^*\left(-\omega\right) a_{\mathrm{in},\omega} + \chi_\mathrm{d}^*\left(-\omega\right) a_{\mathrm{in},-\omega}^\dagger + \chi_\mathrm{x}^\mathrm{TLS*} \left(-\omega\right) a_{\mathrm{in},\omega}^\mathrm{TLS} + \chi_\mathrm{d}^\mathrm{TLS*} \left(-\omega\right) a_{\mathrm{in},-\omega}^{\mathrm{TLS}\dagger} , \label{eq:S10}
\end{align}
\end{subequations}
where
\begin{subequations}
	\begin{align}
	\chi_\mathrm{d}\left(\omega\right) &= \sqrt{\kappa} C (AC -\left|B\right|^2)^{-1}, \\ \chi_\mathrm{x}\left(\omega\right) &= -\sqrt{\kappa} B (AC -\left|B\right|^2)^{-1}, \\ \chi_\mathrm{d}^\mathrm{TLS}\left(\omega\right) &= 2\sqrt{\kappa_N} \alpha^* C (AC -\left|B\right|^2)^{-1} , \\
	\chi_\mathrm{x}^\mathrm{TLS}\left(\omega\right) &= - 2 \sqrt{\kappa_N} \alpha B (AC -\left|B\right|^2)^{-1} .
	\end{align}
\end{subequations}

\begin{figure}
	\centering
	\includegraphics[width=1\textwidth]{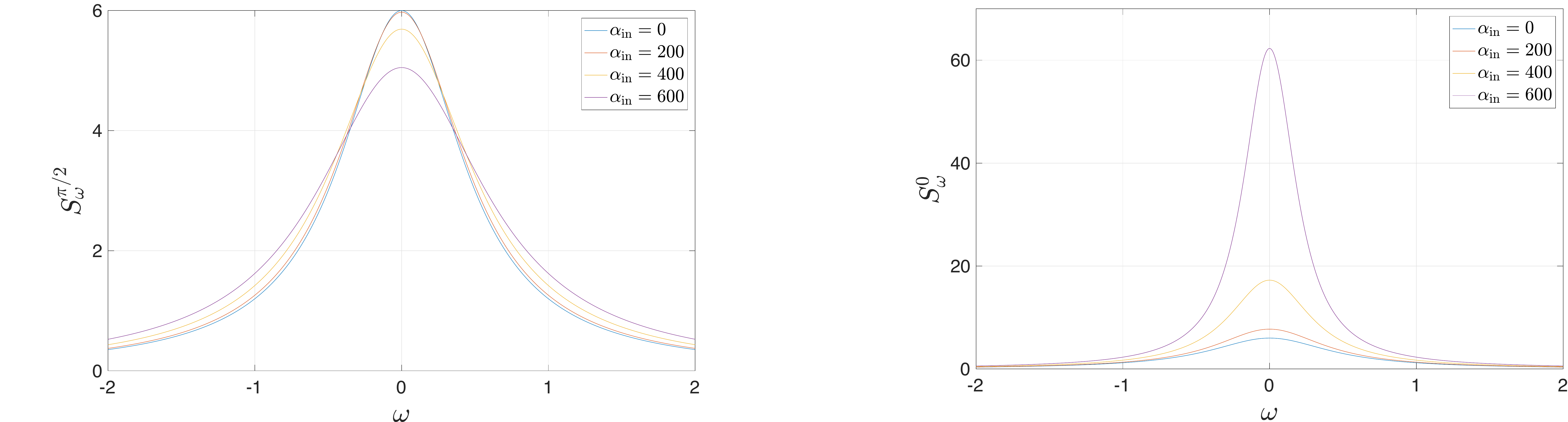}
	\caption{The cavity spectra related to the Holstein-Primakoff couplings
          (a) HP$_-$ and (b) HP$_+$ for the largest uncertainty quadrature
          ($\theta = \pi/2$ and $\theta = 0$, respectively). In (a) the
          linewidth widens as $\alpha_\mathrm{in}$ becomes larger, whereas in
          (b) the linewidth becomes narrower. Here the thermal populations of
          the bosonic and TLS baths are
          $n_\mathrm{th} = n_\mathrm{th}^\mathrm{TLS} = 1$, and in the units of
          $\kappa$, the other parameters are $\Delta = 20$ and
          $\kappa_N = 1.5 \times 10^{-4}$.}\label{fig:spectrum}
\end{figure}

If we assume that the thermal populations of the baths are given by
$\Braket{a_{\mathrm{in},\omega} a_{\mathrm{in},\omega '}^\dagger} =
\left(n_\mathrm{th} + 1\right) \delta \left(\omega - \omega '\right)$ and
$\Braket{a_{\mathrm{in},\omega}^\mathrm{TLS} a_{\mathrm{in},\omega
    '}^{\mathrm{TLS}\dagger}} = \left(n_\mathrm{th}^\mathrm{TLS} + 1\right)
\delta \left(\omega - \omega '\right)$, the cavity spectrum can be written as
\begin{equation}
\label{eq:S14}
\begin{split}
	S_\omega^\theta =&\, \frac{1}{4} \left[ \left( \left|\chi_\mathrm{d}\left(\omega\right)\right|^2 + \left|\chi_\mathrm{x}\left(-\omega\right)\right|^2 \right) \Braket{\left\{a_{\mathrm{in},\omega}, a_{\mathrm{in},\omega}^\dagger\right\}} + \left( \left|\chi_\mathrm{d}\left(-\omega\right)\right|^2 + \left|\chi_\mathrm{x}\left(\omega\right)\right|^2 \right) \Braket{\left\{a_{\mathrm{in},-\omega}^\dagger, a_{\mathrm{in},-\omega}\right\}} \right] \\
	&+ \frac{1}{4} \Big[ \left(\chi_{\mathrm{d}}\left(\omega\right) \chi_{\mathrm{x}}\left(-\omega\right) e^{-i2\theta} + \chi_{\mathrm{d}}^* \left(\omega\right) \chi_{\mathrm{x}}^* \left(-\omega\right) e^{i2\theta}\right) \Braket{\left\{a_{\mathrm{in},\omega}, a_{\mathrm{in},\omega}^\dagger\right\}} \\
	&\qquad + \left(\chi_{\mathrm{d}}\left(-\omega\right) \chi_{\mathrm{x}}\left(\omega\right) e^{-i2\theta} + \chi_{\mathrm{d}}^* \left(-\omega\right) \chi_{\mathrm{x}}^* \left(\omega\right) e^{i2\theta} \right) \Braket{\left\{a_{\mathrm{in},-\omega}^\dagger, a_{\mathrm{in},-\omega}\right\}} \Big] \\
	&+ \frac{1}{4} \Big[ \left( \left|\chi_\mathrm{d}^\mathrm{TLS}\left(\omega\right)\right|^2 + \left|\chi_\mathrm{x}^\mathrm{TLS}\left(-\omega\right)\right|^2 \right) \Braket{\left\{a_{\mathrm{in},\omega}^\mathrm{TLS}, a_{\mathrm{in},\omega}^{\mathrm{TLS}\dagger}\right\}} \\
	&\qquad + \left( \left|\chi_\mathrm{d}^\mathrm{TLS} \left(-\omega\right)\right|^2 +  \left|\chi_\mathrm{x}^\mathrm{TLS}\left(\omega\right)\right|^2 \right) \Braket{\left\{a_{\mathrm{in},-\omega}^{\mathrm{TLS}\dagger}, a_{\mathrm{in},-\omega}^\mathrm{TLS}\right\}} \Big] \\
	&+ \frac{1}{4} \Big[ \left(\chi_\mathrm{d}^\mathrm{TLS}\left(\omega\right) \chi_\mathrm{x}^\mathrm{TLS}\left(-\omega\right) e^{-i2\theta} + \chi_\mathrm{d}^{\mathrm{TLS}*} \left(\omega\right) \chi_\mathrm{x}^{\mathrm{TLS}*} \left(-\omega\right) e^{i2\theta}\right) \Braket{\left\{a_{\mathrm{in},\omega}^\mathrm{TLS}, a_{\mathrm{in},\omega}^{\mathrm{TLS}\dagger}\right\}} \\
	&\qquad + \left(\chi_\mathrm{d}^\mathrm{TLS}\left(-\omega\right) \chi_\mathrm{x}^\mathrm{TLS}\left(\omega\right) e^{-i2\theta} + \chi_\mathrm{d}^{\mathrm{TLS}*} \left(-\omega\right) \chi_\mathrm{x}^{\mathrm{TLS}*} \left(\omega\right) e^{i2\theta} \right) \Braket{\left\{a_{\mathrm{in},-\omega}^{\mathrm{TLS}\dagger}, a_{\mathrm{in},-\omega}^\mathrm{TLS}\right\}} \Big] \\
	=&\, \frac{1}{2} \Big[ \left|\chi_{\mathrm{d}}\left(\omega\right)\right|^2 + \left|\chi_{\mathrm{d}}\left(-\omega\right)\right|^2 + \left|\chi_{\mathrm{x}}\left(\omega\right)\right|^2 + \left|\chi_{\mathrm{x}}\left(-\omega\right)\right|^2 \\
	&\qquad + 2 \cos\left(\theta + \phi\right) \left| \chi_{\mathrm{d}}\left(\omega\right) \chi_{\mathrm{x}}\left(-\omega\right) + \chi_{\mathrm{d}} \left(-\omega\right) \chi_{\mathrm{x}} \left(\omega\right) \right| \Big] \left(n_\mathrm{th} + \frac{1}{2}\right) \\
	&+ \frac{1}{2} \Big[ \left|\chi_\mathrm{d}^\mathrm{TLS}\left(\omega\right)\right|^2 + \left|\chi_\mathrm{d}^\mathrm{TLS}\left(-\omega\right)\right|^2 + \left|\chi_\mathrm{x}^\mathrm{TLS}\left(\omega\right)\right|^2 + \left|\chi_\mathrm{x}^\mathrm{TLS}\left(-\omega\right)\right|^2 \\
	&\qquad + 2 \cos\left(\theta + \phi^\mathrm{TLS}\right) \left| \chi_\mathrm{d}^\mathrm{TLS}\left(\omega\right) \chi_\mathrm{x}^\mathrm{TLS}\left(-\omega\right) + \chi_\mathrm{d}^\mathrm{TLS} \left(-\omega\right) \chi_\mathrm{x}^\mathrm{TLS} \left(\omega\right) \right| \Big] \left(n_\mathrm{th}^\mathrm{TLS} + \frac{1}{2}\right) ,\\
\end{split}
\end{equation}
where
$\phi^{(\mathrm{TLS})} = \mathrm{Arg} \left[ \chi_{\mathrm{d}}^{(\mathrm{TLS})}
  \left(\omega\right) \chi_{\mathrm{x}}^{(\mathrm{TLS})} \left(-\omega\right) +
  \chi_{\mathrm{d}}^{(\mathrm{TLS})} \left(-\omega\right)
  \chi_{\mathrm{x}}^{(\mathrm{TLS})} \left(\omega\right) \right]$. In
Fig. \ref{fig:spectrum}(a) we have plotted the cavity spectrum for the HP$_-$,
and the spectrum related to HP$_+$ coupling derived from Eq. (8b) is presented
in Fig. \ref{fig:spectrum}(b).

\end{document}